\documentclass[twocolumn,prl,aps,superscriptaddress,showpacs]{revtex4}
\usepackage{epsfig}
\usepackage{amsmath}
\usepackage{amsthm}
\usepackage{amssymb,braket}
\usepackage{verbatim}
\usepackage{bm}
\usepackage{hyperref}
\usepackage[all]{hypcap}
\usepackage{here}
\usepackage{sidecap}
\usepackage{float}

\begin{document}

\title{Experimental observation of hidden Berry curvature in inversion-symmetric bulk 2H-WSe$_2$ }

\author{Soohyun Cho}
\affiliation{Institute of Physics and Applied Physics, Yonsei University, Seoul 03722, Korea}
\affiliation{Center for Correlated Electron Systems, Institute for Basic Science (IBS), Seoul 08826, Republic of Korea}

\author{Jin-Hong Park}
\affiliation{Center for Correlated Electron Systems, Institute for Basic Science (IBS), Seoul 08826, Republic of Korea}

\author{Jisook Hong}
\affiliation{Department of Chemistry, Pohang University of Science and Technology, Pohang 37673, Republic of Korea}

\author{Jongkeun Jung}
\affiliation{Center for Correlated Electron Systems, Institute for Basic Science (IBS), Seoul 08826, Republic of Korea}
\affiliation{Department of Physics and Astronomy, Seoul National University (SNU), Seoul 08826, Republic of Korea}

\author{Beom Seo Kim}
\affiliation{Center for Correlated Electron Systems, Institute for Basic Science (IBS), Seoul 08826, Republic of Korea}
\affiliation{Department of Physics and Astronomy, Seoul National University (SNU), Seoul 08826, Republic of Korea}

\author{Garam Han}
\affiliation{Center for Correlated Electron Systems, Institute for Basic Science (IBS), Seoul 08826, Republic of Korea}
\affiliation{Department of Physics and Astronomy, Seoul National University (SNU), Seoul 08826, Republic of Korea}

\author{Wonshik Kyung}
\affiliation{Center for Correlated Electron Systems, Institute for Basic Science (IBS), Seoul 08826, Republic of Korea}
\affiliation{Department of Physics and Astronomy, Seoul National University (SNU), Seoul 08826, Republic of Korea}
\affiliation{Advanced Light Source, Lawrence Berkeley National Laboratory, California 94720, USA}

\author{Yeongkwan Kim}
\affiliation{Department of Physics, Korea Advanced Institute of Science and Technology, Daejeon 34141, Republic of Korea}

\author{S.-K. Mo}
\affiliation{Advanced Light Source, Lawrence Berkeley National Laboratory, California 94720, USA}

\author{J. D. Denlinger}
\affiliation{Advanced Light Source, Lawrence Berkeley National Laboratory, California 94720, USA}

\author{Ji Hoon Shim}
\affiliation{Department of Chemistry, Pohang University of Science and Technology, Pohang 37673, Republic of Korea}
\affiliation{Department of Physics and Division of Advanced Nuclear Engineering, Pohang University of Science and Technology, Pohang 37673, Republic of Korea}

\author{Jung Hoon Han}
\affiliation{Department of Physics, Sungkyunkwan University, Suwon 16419, Republic of Korea}

\author{Changyoung Kim}
\email{changyoung@snu.ac.kr}
\affiliation{Center for Correlated Electron Systems, Institute for Basic Science (IBS), Seoul 08826, Republic of Korea}
\affiliation{Department of Physics and Astronomy, Seoul National University (SNU), Seoul 08826, Republic of Korea}

\author{Seung Ryong Park}
\email{AbePark@inu.ac.kr}
\affiliation{Department of Physics, Incheon National University, Incheon 22012, Republic of Korea}

\date{\today}

\begin{abstract}
We investigate the hidden Berry curvature in bulk 2H-WSe$_2$ by utilizing the surface sensitivity of angle resolved photoemission (ARPES). The symmetry in the electronic structure of transition metal dichalcogenides is used to uniquely determine the local orbital angular momentum (OAM) contribution to the circular dichroism (CD) in ARPES. The extracted CD signals for the $K$ and $K^\prime$ valleys are almost identical but their signs, which should be determined by the valley index, are opposite. In addition, the sign is found to be the same for the two spin-split bands, indicating that it is independent of spin state. These observed CD behaviors are what are expected from Berry curvature of a monolayer of WSe$_2$. In order to see if CD-ARPES is indeed representative of hidden Berry curvature within a layer, we use tight binding analysis as well as density functional calculation to calculate the Berry curvature and local OAM of a monolayer WSe$_2$. We find that measured CD-ARPES is approximately proportional to the calculated Berry curvature as well as local OAM, further supporting our interpretation.
\pacs{79.60.-i, 74.20.Pq, 73.20.-r}
\end{abstract}
\maketitle

The broken inversion symmetry in a monolayer (ML) of transition metal dichalcogenides (TMDCs) 2H-MX$_2$, together with strong spin-orbit coupling (SOC), results in inequivalent valleys with spin splitting at $K$ and $K^\prime$ in the Brillouin zone (BZ) \cite{transistor,dxiao,king,zhu}. These inequivalent valleys at $K$ and $K^\prime$ lead to the valley Hall effect which, unlike the ordinary Hall effect, produces not only charge but also spin imbalance at the edges \cite{dxiao,yao,iwasa,natpho,transistor,jena}. The valley Hall effect has been understood in terms of the Berry curvature \cite{modern,dxiao,natphyrev,gravalley,gravalleyB,jifeng}; the symmetries in 1ML 2H-MX$_2$ causes sign change in the Berry curvature as one goes from one valley ($K$) to an inequivalent valley ($K^\prime$) in the BZ \cite{yao,dxiao,souza,tigh,tight2,jifeng}. This allows us to understand the valley Hall effect in terms of pseudo-spins, and provides possibilities to control the pseudo-spins by an external field \cite{dxiao, conheinz, euse, natrev, jong,imamo,xxu, cui,xxu2,marie}.

On the other hand, the Berry curvature is expected to vanish in the bulk (so does the valley Hall effect) because the bulk TMDCs have an inversion symmetry \cite{yao,cui}. However, one can imagine that the valley Hall in each layer could be non-vanishing - only the sum vanishes. This may naturally introduce the concept of ``hidden Berry curvature'', a non-vanishing Berry curvature localized in each layer. An analogous case can be found in the case of the hidden spin polarization proposed and measured recently \cite{king,schneider,yao,hidden,szhou,spinhid}. Existence of hidden Berry curvature implies that the topology could be determined by local field; the local symmetry determines the physics \cite{hidden,in,cheolhwan}. While experimental verification of hidden Berry phase in the Bloch state is highly desired, standard measurements such as quantum oscillation \cite{sdhberry,berrybise,berryrashba} cannot reveal hidden Berry phase because these measurements represent an averaged quantity, with hidden quantity invisible.

However, if we use a external field \cite{xxu2} or surface sensitive technique such as angle resolved photoemission (ARPES) \cite{hufner}, direct measurement of such hidden Berry curvature may be possible. In fact, the surface sensitivity of ARPES has been utilized recently in the measurement of hidden spin polarization \cite{king,schneider,spinhid,szhou}. Then, the question is if one can measure the Berry curvature by means of ARPES. In this regard, we note a recent proposal that Berry curvature is approximately proportional to the orbital angular momentum (OAM) in the Bloch state \cite{Go}. Since it has been shown that OAM can be measured by circular dichroism (CD) ARPES \cite{seung1,seung2,han,beom,ryu,fengcd,kingcd,jung,kingnat}, one can directly observe the existence of hidden Berry curvature by using CD-ARPES.

In actual measurements, an important challenge lies in the fact that CD-ARPES has contributions other than the one from OAM \cite{cucdnat,cucdprl}. The most notable contribution comes from the geometrical effect which is caused by a mirror symmetry breaking in the experimental geometry. Therefore, how we separate the Berry curvature and geometrical contributions holds the key to successful observation of the hidden Berry curvature. We exploit the unique valley configurations of TMDCs in the BZ to successfully disentangle the two contributions. The observed hidden Berry curvature has opposite signs at $K$ and $K^\prime$ as theoretically predicted. Moreover, we find the hidden Berry curvature exists over a wide range in the BZ. These features are consistently explained within the first principles calculations and tight binding description.

\begin{figure}
\centering \epsfxsize=8.5cm \epsfbox{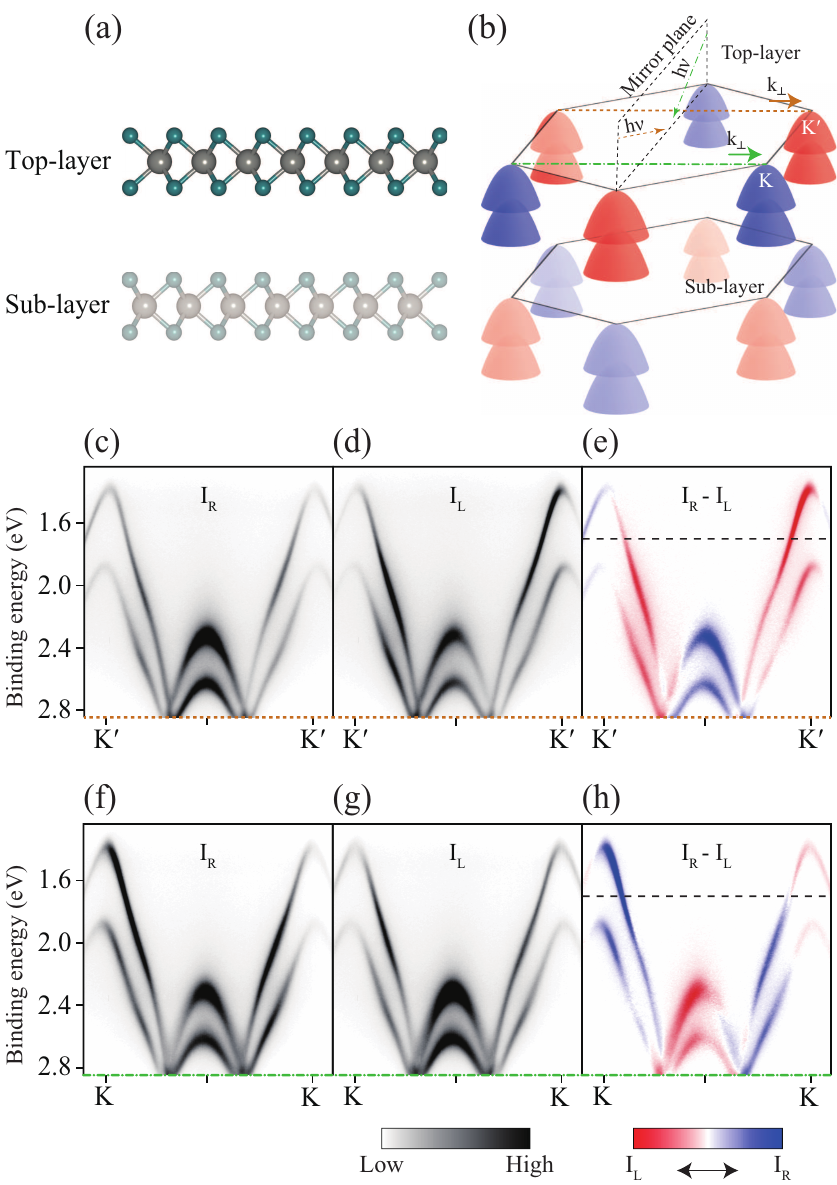} \caption{(Color online). (a) Side view of bulk 2H-WSe$_2$. W and Se atoms are shown as gray and green balls, respectively. (b) Experimental geometry with the hexagonal BZ and spin split bands. $K$ and $K^\prime$ valleys are color coded in blue and red, respectively. The upper and lower BZs represent the top- and sub-layer BZs, respectively. Green dash-dot and brown dashed lines mark the photon incident and cut directions for $K$-$K$ and $K^\prime$-$K^\prime$ cuts, respectively. The same mirror plane was used for the two cases. ARPES data along $K^\prime$ to $K^\prime$ valley taken with (c) right- (RCP) and (d) left- (LCP) circularly polarized 94 eV light. (e) Circular dichroism (CD) obtained from the difference between (c) and (d). This cut corresponds to the brown line in panel (b). (f) RCP, (g) LCP and (h) CD intensities for the $K$ to $K$ cut. (f)-(h) correspond to the cut shown by the green line in panel (b).}
\label{fig1}
\end{figure}

ARPES measurements were performed at the beam line 4.0.3 of the Advanced Light Source at the Lawrence Berkeley National Laboratory. Data were taken with left- and right-circularly polarized (LCP and RCP, respectively) 94 eV light, with the circular polarization of the light better than 80 \%. The energy resolution was better than 20 meV with a momentum resolution of 0.04 {\AA}$^{-1}$. Bulk 2H-WSe$_2$ single crystals were purchased from HQ graphene and were cleaved {\it in situ} at 100 K in a vacuum better than 5$\times$$10^{-11}$ Torr. All the data were taken at 100 K.

Figure 1(a) shows the crystal structure of 2H-WSe$_2$ for which the inversion symmetry is broken for a ML. In the bulk form of 2H-WSe$_2$, the layers are stacked in a way that inversion symmetry is recovered. In the actual experiment, the contribution from the top-layer to the ARPES signal is more than that from the sub-layer, as illustrated by the dimmed color of the sub-layer. Figure 1(b) schematically sketches the electronic structure with the hexagonal BZ of WSe$_2$. The low energy electronic structures of 2H-WSe$_2$ ML was found to be described by the massive Dirac-Fermion model \cite{dxiao,souza,tigh,tight,tight2,beom1}, with hole bands at $K$ and $K^\prime$ points \cite{mo,cvd}. These hole states at $K$ and $K^\prime$ points have local atomic OAM of $2\hbar$ and $-2\hbar$, respectively, which works as the valley index \cite{tigh,tight}. The bands are then spin-split due to the coupling between the spin and OAM.

In the bulk, layers are stacked in a way that $K$ ($K^\prime$) of a layer is at the same momentum position as the $K^\prime$ ($K$) of next layer. Consequently, spin and valley symmetries are restored due to the recovered inversion symmetry and any valley sensitive signal should vanish \cite{cui,tight,yao,zhu}. On the other hand, the in-plane nature of the primary orbital character of the Bloch states ($d_{xy}$, $d_{x^{2}-y^{2}}$, $p_x$ and $p_y$ orbitals) around the $K$ and $K^\prime$ points and the graphene-like phase cancellation as well as the strong spin orbital coupling strongly suppress the interlayer hopping along the $c$-axis and make them quasi-two-dimensional \cite{yao,xcui}. In that case, the valley physics as well as spin-split nature may be retained within each layer as illustrated in Fig. 1(b) by the top- and sub-layer spin split bands (hidden nature). In that case, one may be able to measure the hidden Berry curvature by using ARPES because it preferentially probes the top-layer due to its surface sensitivity as, once again, illustrated by the dimmed color of the sub-layer. Since the signal is preferentially from the top-layer, the situation becomes as if ARPES data is taken from the topmost layer of WSe$_2$, for which the inversion symmetry is broken \cite{king,schneider,szhou,beomseo}.

As mentioned earlier, it was argued that OAM is directly related to the Berry curvature which indeed has opposite signs at the $K$ and $K^\prime$ points as OAM does \cite{tigh,souza}. Then, the hidden Berry curvature may be measured by using CD-ARPES which was shown to be sensitive to OAM. However, CD-ARPES has aforementioned geometrical contribution due to the broken mirror symmetry (chirality) in the experimental geometry. In order to resolve the issue, we exploit the unique character of the electronic structures of TMDCs. The key idea is that, while the contribution from the geometrical effect is an odd function of $k$ about the mirror plane \cite{cucdnat,cucdprl}, we can make the OAM contribution an even function. In that case, the two contributions can be easily isolated from each other. To make the OAM contribution an even function, we use the experimental geometry illustrated in Fig. 1(b). Experimental mirror plane, which is normal to the sample surface and contains the incident light wave vector, is precisely aligned to cross both $K$ and $K^\prime$ points. In such experimental condition, the Berry curvature is mirror symmetric about the experimental mirror plane and so is its contribution to the CD-ARPES. Then, the CD-ARPES is taken along the momentum perpendicular to the mirror plane ($k_\perp$), $i.e.$, from $K$ to $K$ and $K^\prime$ to $K^\prime$ as shown in Fig. 1(b) by green dash-dot and brown dashed lines, respectively. We point out that we kept the same light incident angle for $K$-$K$ and $K^\prime$-$K^\prime$ cuts (note the color pair for the cut and light incidence in Fig. 1(b)) to prevent any contribution other than those from Berry curvature and experimental chirality.

Figures 1(c)-1(e) show data along the $K^\prime$-$K^\prime$ cut. The dispersion is very symmetric with the minimum binding energy at the $K^\prime$ point as expected. However, the intensity varies rather peculiarly; there appears to be no symmetry in the CD intensity in Fig. 1(e). The $K$-$K$ cut in Figs. 1(f)-1(h) shows a similar behavior. While the dispersion is symmetric (and also identical to the $K^\prime$-$K^\prime$ cut), the CD intensity in Fig. 1(h) at a glance does not seem to show a symmetric behavior. However, upon a close look of the CD data in Figs. 1(e) and 1(h), one finds that the two are remarkably similar; the two are almost exact mirror images of each other if the colors are swapped in one of the images. This is already an indication that the CD data reflect certain aspects of the electronic structure that are opposite at the $K$ and $K^\prime$ points, most likely the hidden Berry curvature of bulk 2H-WSe$_2$.

\begin{figure}
\centering \epsfxsize=8.5cm \epsfbox{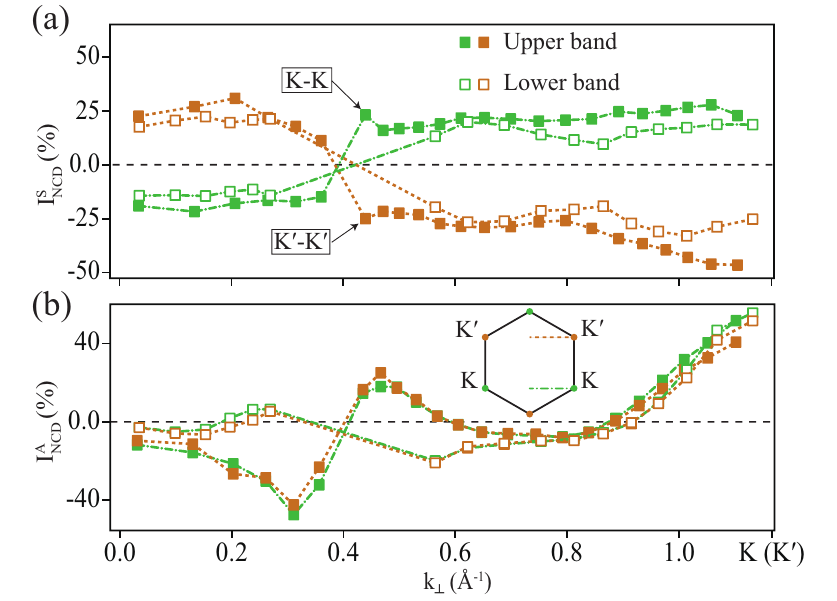} \caption{(Color online) (a) Symmetric ($I^{S}_{NCD}$) and (b) anti-symmetric ($I^{A}_{NCD}$) parts of the normalized CD, $I_{NCD}$, plotted against the momentum of the MDC peak. The plot is made for only half of the momentum range of a cut, between the mirror plane and $K$ or $K^\prime$ point. $I^{S}_{NCD}$ and $I^{A}_{NCD}$ are even and odd functions of $k_\perp$. The filled (empty) squares indicate the upper (lower) spin-split bands.}
\label{fig2}
\end{figure}

We still need more analysis to extract the contribution from the hidden Berry curvature since CD-ARPES has component from the experimental chirality. In order to extract the hidden Berry curvature contribution, we exploit the fact that the contributions from the hidden Berry curvature and photoemission chirality to the CD-ARPES have even and odd parities, respectively. We first define the CD-ARPES as  $I_{CD}=I_R-I_L$ where $I_R$ and $I_L$ are ARPES intensities taken with RCP and LCP, respectively. Even and odd components of $I_{CD}(k)$ are then easily extracted by $I^{S}_{CD}(k)=[I_{CD}(k)+I_{CD}(-k)]/2$ and $I^{A}_{CD}=[I_{CD}(k)-I_{CD}(-k)]/2$ for a momentum distribution curve (MDC) at a binding energy, respectively (see Supplementary Information for details). Remarkably, while the symmetric components are found to be almost exactly opposite for $K$-$K$ and $K^\prime$-$K^\prime$ cuts, the anti-symmetric components are almost identical. The former is what is expected from the Berry curvatures at $K$ and $K^\prime$.

One may use normalized CD $I_{NCD}$ as a quantitative measure of the CD. It is defined as the difference between areas of an MDC peak taken with RCP and LCP, normalized by the sum of them \cite{beom,cucdnat,cucdprl,kingcd,kingnat,ryu,seung1}. The symmetric ($I^{S}_{NCD}$) and antisymmetric ($I^{A}_{NCD}$) components of $I_{NCD}$ can be obtained similarly. $I^{S}_{NCD}$ and $I^{A}_{NCD}$ for all binding energies are plotted in Fig. 2 as a function of the momentum. Effectively, we move along a band and plot symmetric ($I^{S}_{NCD}$) and antisymmetric ($I^{A}_{NCD}$) contributions to the CD against the momentum for each point on the band. There are several aspects to be noticed from the figure. First of all, the symmetric part $I^{S}_{NCD}$ for $K$-$K$ (green) and $K^\prime$-$K^\prime$ (brown) cuts have almost identical behavior except their signs are reversed as seen in Fig. 2(a). As the momentum changes away from $K$ ($K^\prime$), $I^{S}_{NCD}$ maintains the same sign until it changes the sign near $k_\perp\approx 0.4\AA$. As we will show later, this sign changes occurs precisely on the entire $\Gamma$-M line. In addition, we find that $I^{S}_{NCD}$ is almost the same for the two spin-split bands (filled and empty symbols). These observations on the behavior of $I^{S}_{NCD}$ are consistent with what we expect from the Berry curvature; it is valley dependent but independent of the spin-split bands \cite{yao,dxiao,souza, tigh,tight2,jifeng}. On the other hand, the anti-symmetric parts $I^{A}_{NCD}$ from $K$-$K$ and $K^\prime$-$K^\prime$ cuts shown in Fig. 2(b) stay very similar to each other over the whole momentum range. The results indicate careful execution of our experiments and trustworthiness of our analysis.

\begin{figure}
\centering \epsfxsize=8.5cm \epsfbox{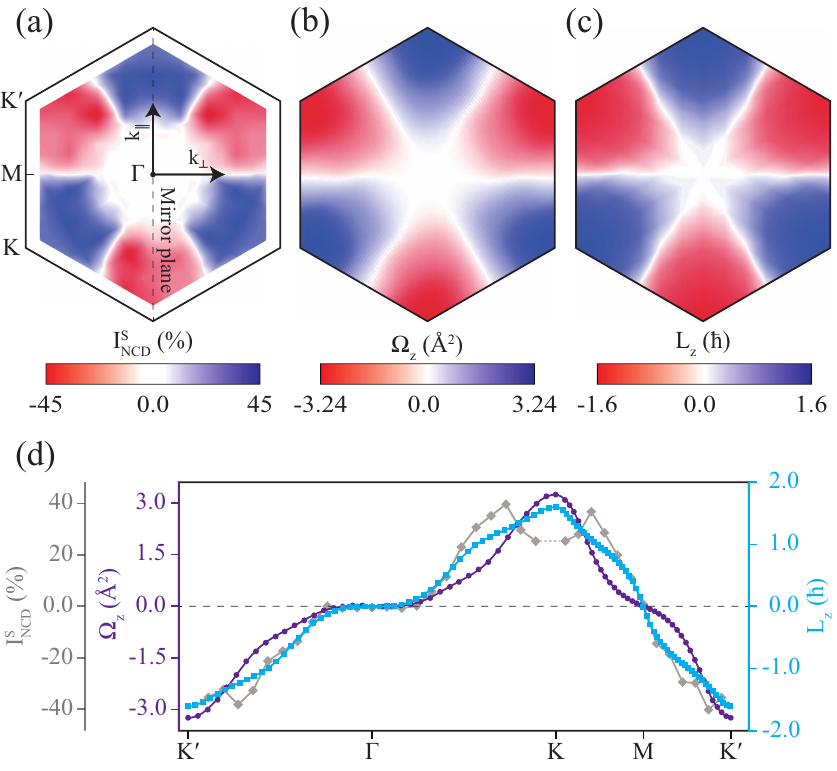} \caption{(Color online) Maps of (a) experimental $I^{S}_{NCD}$, (b) calculated Berry curvature from tight binding model calculation and (c) orbital angular momentum from the first principles calculation. For the map in (a), $I^{S}_{NCD}$ is obtained from the upper valence band CD-ARPES data. The original data covers one third of the BZ data which is then symmetrized to cover the whole BZ. (d) $I^{S}_{NCD}$ (diamond shape), Berry curvature (circle), and OAM (square) values along the $K^\prime$-$\Gamma$-$K$-$M$-$K^\prime$.}
\label{fig3}
\end{figure}

To study the behavior of $I^{S}_{NCD}$ better, we expand the range from a cut to a map of $I^{S}_{NCD}$ that covers the BZ. $I^{S}_{NCD}$ of the upper spin-split band is obtained from the CD-ARPES data and plotted in Fig. 3(a). In addition, in order to understand the connection between $I^{S}_{NCD}$ and Berry curvature as well as OAM, we performed tight binding (TB) analysis and first principles calculation for a ML WSe$_2$. For the Berry curvature calculation, we consider the tight binding Hamiltonian based on the hybridization between W $d$ orbital and Se $p$ orbital (see Supplementary Information for details). In the calculation, the parameters are adjusted until the dispersion fits the experimental one \cite{mo,cvd} and previous TB result \cite{tigh,tight,tight2}. Then, the Berry curvature of the upper band is calculated based on the TKNN formula and its map is plotted in Fig. 3(b). The momentum dependent local OAM ($L_z$) is obtained by density functional theory calculation. The resulting $L_z$ map is depicted in Fig. 3(c). The in-plane components of the Berry curvature and OAM are also calculated but are found to be negligible over the whole BZ and thus are not presented. One can immediately note that the three plots of experimentally obtained $I^{S}_{NCD}$, Berry curvature from TB analysis, and local $L_z$ from DFT calculation show remarkably similar behavior; their signs are determined by the valley indices and change only across the $\Gamma$-M line. In addition, all of them retain significant values quite far away from the $K$ and $K^\prime$ points. Our observation suggests a close relationship between $I^{S}_{NCD}$, Berry curvature and local OAM and thus calls for theoretical exploration of the inter-connectedness between them in TMDC and many other two-dimensional spin-orbit-coupled materials

For a more quantitative comparison, we plot $I^{S}_{NCD}$, Berry curvature and OAM along the high symmetry lines ($K^\prime$-$\Gamma$-$K$-$M$-$K^\prime$). Once again, $I^{S}_{NCD}$, Berry curvature and OAM show very similar behavior. As the Bloch states at the $\Gamma$ and $M$ points possess inversion symmetry, $I^{S}_{NCD}$ and Berry curvature as well as OAM are all zero. One particular aspect worth noting is their behavior near the $\Gamma$ point. They are approximately zero near the $\Gamma$ point but suddenly increase about a third way to $K$ or $K^\prime$ point. Orbital projected band structure from TB calculation shows that this is when the orbital character of the wave function switches from out-of-plane $d_{z^{2}}$ and $p_{z}$ orbitals to in-plane $d_{xy}$, $d_{x^{2}-y^{2}}$, $p_x$ and $p_y$ orbitals. This behavior can be understood from the fact that the local OAM (or valley) is formed by in-plane orbitals. These results strongly suggest that $I^{S}_{NCD}$ is indeed representative of the (hidden) Berry curvature and that the Berry curvature is closely related to the local OAM, at least for TMDCs.

Characteristics of electron wave functions in the momentum space often play very important roles in macroscopic properties of solids. For example, topological nature of an insulator is determined by the characteristics of electron wave function at high symmetric points in the momentum space \cite{seung1,jung}. The Berry curvature which is also embedded in the nature of the electron wave function in the momentum space determines the Berry phase and thus macroscopic properties such as spin and valley Hall effects. Through our work, we demonstrated a way to map out the Berry curvature distribution over the Brillouin zone and provide a direct probe of the topological character of strongly spin-orbit-coupled materials. This stands in contrast with transport measurement of spin and charge which reflect the global momentum-space average of the Berry curvature. In this regards, CD-ARPES can be a useful experimental tool to investigate certain aspects of the phase in electron wave functions \cite{gracd,seung1,seung2,han,beom,ryu,fengcd,kingcd,jung,kingnat} if one can disentangle different contributions in the CD-ARPES.

\acknowledgments
This work was supported by the research program of Institute for Basic Science (Grant No. IBS-R009-G2). S. R. P. acknowledges support from the National Research Foundation of Korea (NRF) (Grant No. 2014R1A1A1002440). The Advanced Light Source is supported by the Office of Basic Energy Sciences of the US DOE under Contract No. DE-AC02-05CH11231.

\end{document}